\documentclass[12pt,showpacs]{revtex4}
\usepackage{graphicx}

\newcommand{\EBP}{E_{\rm BP}}
\newcommand{\ELDP}{E_{\rm LDP}}
\newcommand{\ELC}{E_{\rm loc}}
\begin{document}


\title{Antiresonance and interaction-induced localization in
spin and qubit chains with defects}

\author{M.I. Dykman}
\email{dykman@pa.msu.edu}
\author{L. F. Santos}
\email{santos@pa.msu.edu}

\affiliation{Department of Physics and Astronomy
and the Institute for Quantum Sciences,
Michigan State University, East Lansing, MI 48824}

\begin{abstract}
We study a spin chain with an anisotropic XXZ coupling in a magnetic
field. Such a chain models several proposed types of quantum
computers. The chain contains a defect with a different on-site
energy. The interaction between excitations leads to the onset of
two-excitation states localized next to the defect. In a resonant
situation scattering of excitations on each other might cause decay of
an excitation localized on the defect. We find that destructive
many-particle quantum interference {\it eliminates} this decay. Numerical
results confirm the analytical predictions.
\end{abstract}

\maketitle

\vskip 1 cm

An important motivation for making a quantum computer as envisaged by
Feynman is to simulate real quantum systems. Disordered systems with
interacting excitations are of particular interest in this
context. An important class of such systems are disordered spin-1/2
systems. The spin-spin interaction leads to a complicated excitation
spectrum, that has been described only in special cases where the
system is integrable \cite{integrable1,integrable2}.

A spin-1/2 system can be modelled by a set of interacting two-level
subsystems, qubits. Controlled qubits, with interqubit coupling that
is not turned on and off, are basic elements of several proposed
implementations of a quantum computer (QC)
\cite{liquid_NMR,Makhlin01,mark,mooij99,van_der_Wiel03,Yamamoto02}.
Qubit energies can be often individually varied. This allows one to
explore resonant situations in a multiply-excited system. In contrast
to naturally occurring many-particle resonances in condensed-matter
systems, in a QC the parameters can be tuned into resonance in a
controlled way.

In this Letter we show that a simple spin chain with a defect displays
new types of localized two-excitation states and destructive
two-excitation interference, i.e., an effect of two-excitation
antiresonance. This is a result of the interplay between disorder and
interaction. In turn, localization of excitations on individual qubits
is important for quantum computing, because the qubits can be then
prepared in desired states, which are not destroyed by other
excitations.

We consider a spin chain in a magnetic field. The defect corresponds
to ``diagonal disorder'': the spin energy on the defect site differs
from that on the host site (for example, the magnetic field on the
defect site differs from the field on the host sites). The exchange
coupling is anisotropic and is described by the XXZ model, which
applies to several proposed quantum computers. The Hamiltonian of the
chain with a defect on site $n_0$ is of the form
\begin{eqnarray}
\label{hamiltonian}
&&H =  \frac{1}{2}\sum\nolimits_n \varepsilon^{(n)} \sigma_{n}^{z} +
\frac{1}{4}\sum\nolimits_n \sum\nolimits_{i=x,y,z}
J_{ii}\sigma_n^i\sigma_{n+1}^i,\\
&& \varepsilon^{(n)}= \varepsilon +
g\delta_{n,n_0},\qquad
J_{xx}=J_{yy}=J,\qquad J_{zz}=J\Delta .\nonumber
\end{eqnarray}
Here, $\sigma_n^i$ are the Pauli matrices and $\varepsilon^{(n)}$ is a
spin-flip energy on site $n$. All $\varepsilon^{(n)}$ are the same
except for the site $n=n_0$. This is not the condition in which a QC
should normally operate, because excitations are mostly delocalized
and are scattered by each other. But even here, as we show, a
multiply-excited system has strongly localized states.

In Eq.~(\ref{hamiltonian}), the defect is characterized by the excess
energy $g$ . The parameter $\Delta$ determines the coupling
anisotropy. We assume that $|\Delta| \gg 1$, as in the case of a QC
based on electrons on helium, for example \cite{mark}. To simplify
notations, we further assume that the spin coupling is
antiferromagnetic: $J,\Delta > 0$.

For large on-site energies $\varepsilon^{(n)}$, the ground state of
the system corresponds to all spins pointing downward,
$\langle \sigma_n^z\rangle = -1$.  The number of excitations is the
number of spins pointing upward. It is conserved, because the total
spin projection $\sum\sigma_n^z$ commutes with $H$. In what follows we
count energy off from the ground-state energy.

Single-spin excitations for the Hamiltonian (\ref{hamiltonian}) are
well understood \cite{koster54}. In an infinite chain they are either
of the spin-wave type  (``magnons'') or localized on the
defect. The magnon energies form a band of width $2J$ centered at
$\varepsilon_1=\varepsilon-J\Delta$. The localized excitation has
energy $\varepsilon_1+(g^2+J^2)^{1/2}\,{\rm sgn}g$. We will be
interested in the case $g\gg J$, when the excitation is
strongly localized on the defect.

Two-spin excitations are well understood for an ideal chain
\cite{Bethe}. For strong anisotropy, $\Delta\gg 1$, they are either
uncoupled magnons or bound pairs (BPs) of flipped neighboring spins
that propagate together. The two-magnon energy band is centered at
$2\varepsilon_1$ and has a width of $4J$. When excitations are on
neighboring sites, their energy is changed by $J\Delta\gg J$, because
of the term $ (J\Delta/4)\sum_n\sigma_n^z\sigma_{n+1}^z$ in $H$
(\ref{hamiltonian}). Respectively, the BP band is centered at
$\EBP^{(0)}\approx 2\varepsilon_1+J\Delta$ and is well separated from
the two-magnon band.  Motion of a BP occurs via virtual
dissociation, where one excitation hops to a neighboring nonexcited
site and then the second excitation moves next to it. The bandwidth
resulting from this process can be calculated by perturbation theory
in $1/\Delta$. It is given by $J/\Delta$ and is parametrically smaller
than the magnon bandwidth.

In the presence of a defect, there emerges an additional
two-excitation state where one excitation is localized on the defect
and the other is in an extended (magnon-type) state. We call it a
localized-delocalized pair (LDP). The LDP band is centered at
$\ELDP^{(0)}\approx 2\varepsilon_1 + g$ and has a width of the
one-magnon band $2J$ (see below).

If the LDP energy band overlaps with the BP band, one would
expect that the two types of excitations mix together. In other words,
the propagating magnon of an LDP can scatter off the localized
excitation, and they together would move away from the defect as a
BP. This would lead to magnon-induced decay of a localized
excitation. We show below that the decay does not happen because of
destructive quantum interference.

The wave function of a chain with two excitations is a linear
superposition $\psi_2 = \sum_{n<m} a(n,m) \phi (n,m)$, where
$\phi(n,m)$ is the state where the spins on sites $n$ and $m$ are
pointing upward and all other spins are pointing downward. From
Eq.~(\ref{hamiltonian}), the Schr\"odinger equation for the
coefficients $a(n,m)$ has the form
\begin{eqnarray}
\left( 2{\varepsilon}_{1} + g \delta_{n,n_0}+ g \delta_{m,n_0}
+J\Delta  \delta_{m,n+1}\right)a(n,m)\nonumber\\
+(J/2)\sum\nolimits_{k=\pm 1}[a(n+k,m) + a(n,m+k)]=E_2a(n,m).
\label{a(n,m)}
\end{eqnarray}
Here, $E_2$ is the energy of a two-excitation state counted off from
the ground-state energy, and $a(n,n)\equiv 0$ (two excitations may not
be placed on one site).

The defect may lead to a localized state of a bound pair with energy
close to the energy of a propagating pair $\EBP^{(0)}$. This state is
centered on sites $(n_0+1,n_0+2)$, or equivalently $(n_0-1,n_0-2)$. To
show this we seek the major term in the solution of Eq.~(\ref{a(n,m)})
in the form
\begin{eqnarray}
\label{major}
a^{(0)}(n,m)=
A\delta_{n,n_0+1}\delta_{m,n_0+2}+
C_1e^{i\theta_1(n-n_0-2)}\delta_{m,n+1}\Theta(n-n_0-2)\nonumber\\
+
\left(C_2e^{i\theta_2(m-n_0-2)}
+C_2'e^{-i\theta_2(m-n_0-2)}\right)\delta_{n,n_0}\Theta(m-n_0-2).
\end{eqnarray}
Here, $A$ is the amplitude of the pair on sites $(n_0+1,n_0+2)$, $C_1$
is the amplitude of a BP propagating away from the defect with wave
number $\theta_1$, and $C_2$ and $C_2'$ are the amplitudes of LDP
waves in which one excitation is on the defect whereas the other is
propagating away from and toward the defect, respectively, with wave
number $\theta_2$; $\Theta(m)$ is the step function, it is equal to
$1$ for $m\geq 0$ and 0 for $m<0$. For concreteness, we consider
excitations ``to the right'' from the defect, $m>n_0$. They do not mix
with the excitations ``to the left'', to first order in $\Delta^{-1}$.

The BP localized next to the defect would be described by the wave
function (\ref{major}) with $C_2'=0$ and with Im~$\theta_{1,2}>0$. On
the other hand,  resonant mixing of LDPs and BPs would be
described as scattering of an LDP wave incident on the defect
($C_2'=1, \theta_2>0$) into a reflected LDP with amplitude $C_2$ and
a BP with amplitude $C_1$ and $\theta_1>0$.

Eqs.~(\ref{a(n,m)}), (\ref{major}) with $n=m-1>n_0+2$ and
$n=n_0,m>n_0+3$ give, respectively, the dispersion laws of BPs and
LDPs far from the defect. From Eq.~(\ref{a(n,m)}), a BP on sites
$(n,n+1)$ is directly coupled to nonresonant dissociated pairs on
sites $(n-1,n+1)$ and $(n,n+2)$. Similarly, an LDP on sites $(n_0,m)$
is coupled to resonant pairs on sites $(n_0,m\pm 1)$ and to
nonresonant pairs on sites $(n_0\pm 1,m)$. Finding the amplitudes of
the nonresonant pairs to first order in $\Delta^{-1}$ from
Eq.~(\ref{a(n,m)}) and substituting them back into the equations for
$C_1$ and $C_2$, we obtain, respectively, the BP and LDP dispersion
laws:
\begin{eqnarray}
\label{dispersion}
\EBP(\theta)=\EBP^{(0)}+(J/2\Delta)\cos\theta,\qquad
E_{\rm BP}^{(0)}= 2\varepsilon_1 + J\Delta + (J/2\Delta),\\
\ELDP(\theta)=\ELDP^{(0)}+J\cos\theta,\qquad
\ELDP^{(0)}=2\varepsilon_1 + g + (J/2\Delta)\nonumber
\end{eqnarray}
(the renormalization of the energy spectrum is calculated for $|E_2-
\EBP^{(0)}|\lesssim J$).

Wave mixing near the defect can be described in a similar way by
writing Eqs.~(\ref{a(n,m)}), (\ref{major}) for the sites
$(n_0+2,n_0+3),(n_0+1, n_0+2)$, and $(n_0,n_0+2)$ and eliminating
nonresonant pair states by perturbation theory. For energies $E_2$
close to $\EBP^{(0)}$ this gives
\begin{equation}
\label{coupled}
C_1e^{-i\theta_1}=A+C_2e^{i\theta_2}+
C_2'e^{-i\theta_2}, \qquad
A=C_2e^{-i\theta_2}+
C_2'e^{i\theta_2}+\Delta^{-1}(C_2+C_2'),
\end{equation}
\begin{eqnarray}
\label{coupled_2}
\left[4(J\Delta-g-J\cos\theta_2)-J\Delta^{-1}\right]A +
2J(C_2+C_2')\nonumber\\
+J\Delta^{-1}(C_2e^{i\theta_2}+C_2'e^{-i\theta_2})
 + J\Delta^{-1}C_1=0
\end{eqnarray}
(the corrections $\sim \Delta^{-1}$ to the first equation
(\ref{coupled}) come from higher order terms of the perturbation
theory; they are not important for the present calculation).

Eqs.~(\ref{coupled}), (\ref{coupled_2}) contain two dimensionless
parameters, $\Delta$ and $\alpha=(J\Delta-g)/J$, which are related to
the three energy scales in the system: the distance between the BP and LDP
bands $J\Delta - g$ and the LDP and BP bandwidths $J$ and
$J/\Delta$.

To find the state localized on sites $(n_0+1,n_0+2)$, we set
$C_2'=0$. Then, for $\Delta \gg 1$ the solution of
Eqs.~(\ref{coupled}), (\ref{coupled_2}) can be approximated by
$\exp(-i\theta_2)\approx 2J^{-1}(J\Delta -g)
+(2\Delta)^{-1}[\exp(i\theta_1) -1]$ (the term $\propto \Delta^{-1}$
is important only for $|\alpha|\gg 1$ and should be modified for
$|\alpha|\sim 1$). The energy of the localized state $E_{\rm loc}$ and
the reciprocal localization lengths Im~$\theta_{1,2}$ are obtained by
substituting this expression into the equation $E_{\rm
loc}=\EBP(\theta_1)=\ELDP(\theta_2)$.

For $|\alpha|\gg 1$ the BP and LDP bands are far from
each other, and
\begin{equation}
\label{nonresonant_1}
E_{\rm loc}- \EBP^{(0)}\approx {J\over 4\Delta}{\Delta^2-2\alpha\Delta
+ 2\alpha^2\over \alpha(\Delta - \alpha)}, \qquad \alpha={J\Delta - g\over J}.
\end{equation}
The localized state amplitude is $A$. The state is mostly hybridized
with the BP band, with $C_1/A\approx \alpha/(\Delta - \alpha)$ and
Im~$\theta_1=\ln[(\Delta - \alpha)/\alpha]$. The admixture of the LDP
band is small, $C_2/A\sim 1/2\alpha \ll 1$. In a way, the localized
state is a ``surface'' state of the BP band, with the role of the
surface being played by the defect. It emerges for $\alpha<\Delta/2$.

As the BP and LDP bands approach each other and $\alpha$ decreases,
the energy $\ELC$ moves away from the BP band. The localized state
becomes stronger hybridized with the LDP band than with the BP band,
with $C_2/A\approx 1/2\alpha$ and $C_1/A\sim \alpha/\Delta \ll 1$ for
$|\alpha|\ll \Delta$. The energy separation from the LDP band is
\begin{equation}
\label{nonresonant_2}
E_{\rm loc}-\ELDP^{(0)}\approx J[\alpha+(4\alpha)^{-1}] \qquad (\alpha\sim 1).
\end{equation}
It is of the order of the LDP bandwidth $J$. The localized state
exists for $|\alpha|> 1/2$. The reciprocal localization length
Im~$\theta_2=\ln (2|\alpha|)$ goes to zero as $|\alpha|\to 1/2$. As
the BP band goes through the LDP band with varying $J\Delta -g$, the
localized state first merges with the LDP band and disappears for
$\alpha = 1/2$, and then is split off from the opposite side of the
LDP band for $\alpha=-1/2$. In more details the dependence of $\ELC$
on parameters is discussed elsewhere \cite{us_archive}, along with the
onset of other localized two-excitation states.

We now consider the possibility of resonant mixing of LDP and BP
states where the bands overlap. As mentioned above, we can do it by
studying scattering of an LDP magnon off the excitation on the defect,
i.e., by setting $C_2'=1$ in Eqs.~(\ref{coupled}), (\ref{coupled_2}).

The LDP band is much broader than the BP band. From
Eqs.~(\ref{dispersion}), LDPs with energies within the BP band have
wave numbers $\theta_2$ such that $\cos\theta_2\approx (J\Delta -
g)/J$. For such LDPs, $C_2\approx -C_2'=-1$ from
Eq.~(\ref{coupled_2}), i.e., they are fully reflected from the
defect. The amplitude of the BP wave is $C_1=0$, to zeroth order in
$\Delta^{-1}$. This happens even though the BP on sites
$(n_0+1,n_0+2)$ has a large amplitude $A\approx 2i\sin\theta_2$.

\begin{figure}[ht]
\includegraphics[width=5.2in]{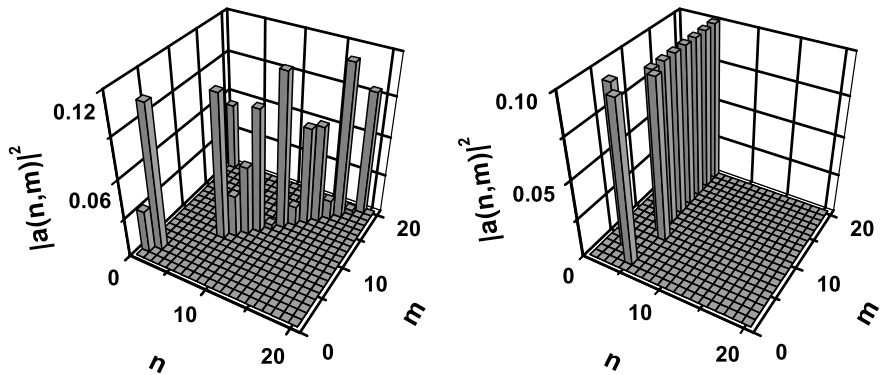}
\noindent
\caption{Amplitudes of two-excitation states on sites $(n,m)$ in a
periodic chain. Left and right panels show, respectively, bound pairs
and localized-delocalized pairs with nearly same energy. The chain
length is 20, the defect is on site $n_0=5$, and $g/J=\Delta =
10$. The BP is located on sites $m=n+1$ and is well approximated by
$a(n,n+1)\propto \sin[10\pi (n-6)/17]$. Because of the antiresonance,
the BP amplitude is equal to zero on sites $n_0, n_0\pm 1, n_0\pm
2$. The LDP is located on sites $(n_0, m)$, with $m-n_0>1({\rm
mod}\,20)$, and is well described by $a(n_0,m)\propto
\sin[(m-1)\pi/2]$. The LDP is strongly hybridized with the pairs on
sites $(n_0+k,n_0+2k)\; (k=\pm 1)$, which have same amplitude, as
expected. }
\label{fig:histograms}
\end{figure}

The vanishing of the BP wave is a result of destructive interference,
or antiresonance, as seen from the first of equations
(\ref{coupled}). The BP wave is coupled to both the pair
$(n_0+1,n_0+2)$ and the pair on sites $(n_0,n_0+3)$, which has an
amplitude $C_2\exp(i\theta_2)+C_2'\exp(-i\theta_2)$. For resonant
values of $\theta_2$ the complex amplitudes of these two pairs are
equal and opposite in sign, they cancel each
other. One can show in a similar way that if a BP wave is incident on
a defect, it will be fully reflected and no LDP waves will be excited.
Absence of mixing of BP and LDP states at resonance is illustrated in
Fig.~\ref{fig:histograms}, which is obtained by direct diagonalization
of Eq.~(\ref{a(n,m)}).

In this paper the effect of antiresonance has been found for
many-particle excitations. It leads to localization of an excitation
on a defect in an XXZ spin or qubit chain even where the excitation
would be expected to decay via multiparticle scattering. We also found
new localized two-excitation states. Their energy and localization
length depend on the interrelation between the interaction anisotropy
$\Delta^{-1}$ and the relative excess energy of an excitation on the
defect site $g/J$. The results indicate that quantum computers can be
advantageous for studying new many-body effects in systems with
disorder.

From the standpoint of quantum computing, the many-particle
localization that we discuss is important, because it is much easier
to manipulate excitations when they are localized on individual
qubits. Of interest for quantum computing is also the occurrence of
different types of localized two-excitation states on neighboring
qubits. It allows creating on these qubits entangled Bell-type states,
which are to some extent similar to recently studied entangled
excitonic states in a quantum dot \cite{carlo}, but which appear in
the present case in an extended system. We expect that coherent
multi-qubit states can be created in a chain with a defect as well.

\begin{acknowledgments}
This research was supported in part by the NSF through grant
No. ITR-0085922 and by the Institute for Quantum Sciences at MSU.
\end{acknowledgments}


\end{document}